\documentclass[aps, prd, nofootinbib, preprint, superscriptaddress]{revtex4-1}

\usepackage[top = 1in, bottom = 1in, left = 1in, right = 1in]{geometry} 

\usepackage{setspace}
\usepackage{amsmath}
\usepackage{mathtools}
\usepackage{amssymb}
\usepackage{amsthm}
\usepackage{thmtools}
\usepackage{thm-restate}
\usepackage{hyperref}
\hypersetup{
    colorlinks=true,
    linkcolor=blue,
    citecolor=magenta,
    urlcolor=cyan
} 
\usepackage{complexity}
\usepackage{enumerate}
\usepackage{todonotes}
\usepackage{algorithm}
\usepackage{comment}
\usepackage{algpseudocode} 

\usepackage{tikz}
\usetikzlibrary{decorations.pathmorphing}
\usetikzlibrary{arrows,positioning,shapes.geometric}
\usepackage{quantikz}

\newcommand{\RR}{\mathbb{R}} 
\newcommand{\CC}{\mathbb{C}} 

\renewcommand{\ket}[1]{| #1 \rangle} 		
\renewcommand{\bra}[1]{\langle #1 |} 		
\renewcommand{\braket}[2]{\langle #1 | #2 \rangle}

\newcommand{\SN}{\mathrm{SN}}

\newtheorem{theorem}{Theorem}
\newtheorem*{thesis}{Physical Extended Church--Turing Thesis}

\renewcommand{\d}{\mathrm{d}}		
\newcommand{\eref}[1]{Eq.~(\ref{#1})}		

\usepackage{xcolor}

\begin{document}
\count\footins = 1000 
\title{Semiclassical Gravity Efficiently Solves $\NP$-Complete Problems}
\author{Matthew Fox}
\email{matthew.fox@colorado.edu}
\address{Department of Physics, University of Colorado, Boulder, CO 80309, USA}
\author{Chaitanya Karamchedu}
\email{cdkaram@umd.edu}
\address{Department of Computer Science, University of Maryland, College Park, MD 20742, USA}
\author{Sotirios Mygdalas}
\email{smygdalas@pitp.ca}
\thanks{\\All authors contributed equally.}
\address{Perimeter Institute for Theoretical Physics, Waterloo, ON N2L 2Y5, Canada}
\address{Department of Physics \& Astronomy, University of Waterloo, Waterloo, ON N2L 3W8, Canada}
\date{\today}

\begin{abstract}
Assuming the gravitational field is classical and that it couples to quantum fields via the semiclassical Einstein field equations, we show that the weak-field dynamics of a massive and non-relativistic qubit can in principle be used to solve an $\NP$-complete problem in polynomial time. We attribute this vast computational power to the non-linear dynamics afforded by the semiclassical Einstein field equations. Consequently, the above two assumptions entail a violation of the Physical Extended Church--Turing Thesis, which we regard as evidence for the quantization of gravity.
\end{abstract}

\maketitle

\section{Introduction}

The road to quantum gravity is not easy. Stuck in our armchairs with little to no experimental guidance, it is prudent to ponder why we think gravity is quantum-mechanical at all. Could it be that the contrary is true, that gravity is not fundamentally quantum but classical? Of course, this question is hardly new. It has a long and rich history, and traces back to the seminal works of M{\o}ller \cite{Lichnerowicz:1962gnh} and Rosenfeld \cite{ROSENFELD1963353} in the 1960s.

Treating gravity classically but matter quantum-mechanically embodies \emph{semiclassical gravity}, which naturally divides into two parts. The first part is the effect of curved spacetimes on quantum fields. This part is well-understood (full books like \cite{Wald:1995yp,Birrell_Davies_1982,Mukhanov:2007zz,Parker:2009uva} have been written on it) and it has proven to be invaluable because of Hawking's discovery that gravitational fields can create particles \cite{hawking1975particle}. The second part is the effect of quantum fields on curved spacetimes. This part is not nearly as understood as the first, largely because it is not clear what it means to superpose spacetimes, which appears to be an inevitability \cite{Wal84}.

Within the quantum gravity community, semiclassical gravity is generally regarded as an approximation to an as yet undiscovered theory of quantum gravity \cite{Wald:1995yp,Birrell_Davies_1982,Mukhanov:2007zz,Parker:2009uva}. However, despite many ongoing experiments (e.g., \cite{MarlettoVidal_GravityMediatedEntanglement_2017, Berglund_2023, Namdar_TripleInterference, tobar2024detecting}), there is no definitive experimental evidence that gravity is quantized. As a result, quantum gravity is supported primarily by theoretical arguments, such as those by Eppley and Hannah \cite{eppley1977necessity} and Page and Geilker \cite{Page_1981}. However, many of these arguments are considered inconclusive \cite{mattingly2005quantum,mattingly2006eppley,Carlip_2008, Grossardt_2022, Bahrami_2014, kent2018AgainstEH}, highlighting the need for a new case for the quantization of gravity.

In this paper, we provide such an argument by exploring the computational consequences of a fundamentally semiclassical world. In particular, we show that if semiclassical gravity (as described by the semiclassical Einstein field equations) is correct, then it is possible to solve $\NP$-complete problems in polynomial time. As this challenges commonly held beliefs about what is and what is not efficiently computable by physical devices (beliefs which are encapsulated in a strong and physical version of the Church--Turing Thesis), our results suggest that semiclassical gravity cannot be a correct description of physical reality.

Our main result hinges on an algorithm due to Bao, Bouland, and Jordan \cite{Bao_2016}, which itself is based on an earlier algorithm by Abrams and Lloyd \cite{Abrams_1998}. Essentially, the Abrams--Lloyd algorithm showed how to use Weinberg's non-linear theory of quantum mechanics \cite{WEINBERG1989336ha} to efficiently solve hard computational problems (e.g., $\NP$-complete problems). Several years later, the Bao--Bouland--Jordan algorithm extended this result and showed that it is, in fact, a more general phenomenon: basically \emph{any} non-linear theory of quantum mechanics can be used to efficiently solve hard computational problems. Our conclusions follow from combining the Bao--Bouland--Jordan algorithm with arguments due to Bahrami et al.~\cite{Bahrami_2014} and Giulini et al.~\cite{Giulini_2023} that the weak-field evolution of a massive qubit in semiclassical gravity is necessarily non-linear. Incidentally, our work is consistent with related work by Kent et al. \cite{Kent_2005,Kent_2021,Fedida_2026}, which we discuss in Section~\ref{sec:discussion}.

\section{$\NP$ and the Physical Extended Church--Turing Thesis}
\label{sec:NPandChurchTuring}

Computational complexity theory seeks to classify problems according to their intrinsic difficulty, often by quantifying the resources (e.g., time or memory) required to solve them. In this paper, we are primarily concerned with how the difficulty of a problem changes when a candidate solution, or \emph{proof}, is provided. Indeed, many problems exhibit an asymmetry in the time required to \emph{find} a solution versus the time required to \emph{verify} a solution. Consider, for example, a Sudoku puzzle: checking a completed grid is far easier than constructing one from scratch. However, it is not known if this distinction reflects a genuine difference in computational difficulty. In computational complexity theory, this question is formalized in the famous $\P$ versus $\NP$ problem, which we now discuss.

Intuitively, the complexity class $\P$ (``polynomial time'') consists of all the computational problems that can be efficiently solved (a.k.a. \emph{decided}) by a classical computer.\footnote{In fact, the set of computational problems that efficient classical computers can solve is not $\P$, but $\BPP$ (``bounded-error probabilistic polynomial time''), which is the probabilistic analogue of $\P$. However, it is generally believed that $\P = \BPP$ \cite{AB09}, so in this discussion we have chosen to only talk about $\P$.} For example, deciding if a binary string is palindromic or not is in $\P$. More formally, let $\{0,1\}^n$ be the set of all $n$-bit binary strings, $\{0,1\}^*$ the set of all finite-length binary strings (i.e., $\{0,1\}^* = \bigcup_{n \geq 1} \{0,1\}^n$), and $|x|$ the length of a string $x$. Then, $\P$ consists of all the functions $L : \{0,1\}^* \rightarrow \{0,1\}$ (a.k.a. \emph{languages}) for which there exists a deterministic and polynomial time classical algorithm $M$ (i.e., a \emph{deterministic and polynomial time Turing machine}) such that for all inputs $x \in \{0,1\}^*$, $L(x) = 1$ if and only if $M(x) = 1$.

On the other hand, the complexity class $\NP$ (``non-deterministic polynomial time") consists of all of the computational problems for which a valid proof can be efficiently checked by a classical computer. For example, deciding if a logical Boolean formula like $(x_1 \lor x_2) \land (x_3 \lor \neg x_2)$, where $x_1, x_2, x_3 \in \{0,1\}$, has a valid assignment is in $\NP$.\footnote{This problem is closely related to the Boolean Satisfiability Problem $\SAT$ \cite{AB09}.} In particular, $L \in \NP$ if and only if there exists a deterministic and polynomial time algorithm $M$ and a polynomial $p$ such that for all inputs $x \in \{0,1\}^*$, $L(x) = 1$ if and only if there exists $y \in \{0,1\}^{p(|x|)}$ (usually called a \emph{proof}, \emph{certificate}, or \emph{witness}) such that $M(x,y) = 1$. In other words, $L \in \NP$ if and only if for all $x \in \{0,1\}^*$, if $L(x) = 1$, then there exists a proof that will cause $M$ to accept $x$, and if $L(x) = 0$, then there is no proof that will cause $M$ to accept $x$.

Evidently, $\P \subseteq \NP$, and it is a major open question (in fact, a Millennium Prize Problem \cite{carlson2023millennium}) whether $\P = \NP$. The general consensus is that $\P \neq \NP$ because the alternative suggests that proving a mathematical statement is just as easy as verifying that a given proof is correct, which is something that mathematical experience says is false \cite{Impagliazzo}.

Now, the entire premise of quantum computation is that it changes the narrative of what can and cannot be computed efficiently in the physical world. In particular, there exist problems widely believed to be in $\NP \backslash \P$ that can be solved efficiently on a quantum computer. The most famous such problem is integer factorization, for which Shor exhibited his eponymous polynomial-time quantum algorithm \cite{Sho99}.

That said, it is generally expected that the hardest problems in $\NP$---the so-called \emph{$\NP$-complete} problems---do not admit efficient quantum algorithms \cite{Nielsen_Chuang_2010, aaronson2005npcompleteproblemsphysicalreality}. In part, this belief stems from the idea that $\NP$-complete problems appear to lack the requisite structure for there to be an exponential quantum speed-up \cite{aaronson2005npcompleteproblemsphysicalreality}. Of course, this is not to say that quantum computers offer no advantage for these problems. Grover's algorithm, for example, solves any $\NP$-complete problem quadratically faster than any classical brute-force algorithm \cite{Grover_1998}. However, this advantage is modest and does not provide the exponential speed-up required to solve an $\NP$-complete problem in polynomial time on a quantum computer. It is also worth noting that if quantum computers cannot solve $\NP$-complete problems in polynomial time, then $\P \neq \NP$ \cite{Nielsen_Chuang_2010}.

More generally, it is a tenet of quantum information theory, and of the theory of computation itself, that no physically reasonable model of computation (perhaps even more powerful than quantum computation) can efficiently solve an $\NP$-complete problem. This expectation is stipulated in the \emph{Physical Extended Church--Turing Thesis}, which was originally proposed by Freedman, Kitaev, Larsen, and Wang in the early 2000s \cite{freedman2002topologicalquantumcomputation} and extended an earlier ``physical'' version of the Church--Turing Thesis by Deutsch in the 1980s \cite{10.1098/rspa.1985.0070}.

\begin{thesis}[PECTT]
No physical procedure can decide an $\NP$-complete problem in polynomially many steps.
\label{thesis:PECTT}
\end{thesis}

While motivated from computer science, this thesis is largely a postulate about the physical world as it articulates what can and cannot be computed efficiently by physical devices. Thus, in the context of demarcating different physical theories, the PECTT dictates that a candidate theory should be deemed implausible if it can be exploited to efficiently solve an $\NP$-complete problem. Ultimately, we will show that semiclassical gravity is such a theory.

\section{Semiclassical Gravity and its Weak-Field Limit}

Historically, the most popular theory of semiclassical gravity is described by the \textit{semiclassical Einstein field equations (semiclassical EFEs)},\footnote{We employ the natural system of units in which $c = G = \hbar = 1$.}
\begin{equation}
R_{\mu\nu} - \frac{1}{2}g_{\mu\nu}R =  8\pi \bra{\Psi} \hat{T}_{\mu \nu} \ket{\Psi},
\label{eq:semiclassicalEFEs}
\end{equation}
where the (unquantized) metric $g_{\mu\nu}$ is sourced by the expectation value of the energy-momentum tensor $\hat{T}_{\mu \nu}$ of quantum matter on spacetime \cite{Giulini_2023}. This theory realizes M{\o}ller's \cite{Lichnerowicz:1962gnh} and Rosenfeld's \cite{ROSENFELD1963353} original proposals and, more formally, is known to arise from an action principle \cite{etde_5892929,Kibble_1980}. Despite several technical concerns (see \cite{Carlip_2008} and references therein), the expectation value $\bra{\Psi}\hat{T}_{\mu \nu}\ket{\Psi}$ can be constructed in a consistent and axiomatic way \cite{Wald:1995yp, Giulini_2023}. Moreover, the semiclassical EFEs are well-defined when the fluctuations of $\hat{T}_{\mu \nu}$ are not very large in the state $\ket{\Psi}$, i.e., when macroscopic energy-momentum configurations are not placed in superpositions \cite{DIOSI1984199,Page_1981}. As we discuss in the next subsection, the weak-field limit of these equations manifest many interesting and testable gravitational effects, thanks to the non-linearity introduced by the semiclassical EFEs.

\subsection{The Weak-Field Limit of Semiclassical Gravity}
\label{sec:weakfieldlimit}

As shown in \cite{Bahrami_2014, Giulini_2023}, in the Newtonian limit (i.e., assuming $g_{\mu \nu} = \eta_{\mu \nu} + h_{\mu \nu}$ with $|h_{\mu\nu}| \ll 1$ and non-relativistic velocities such that the energy density $\hat{T}_{00}$ dominates the rest of $\hat{T}_{\mu\nu}$), \eref{eq:semiclassicalEFEs} reduces to the \emph{Schr\"odinger--Newton (SN) equation},
\begin{equation}
i\, \partial_t \Psi(r,t) = \left(-\frac{1}{2m}\nabla^2 + V_G(\Psi)\right)\Psi(r,t),
\label{eq:SNEquation}
\end{equation}
where
\begin{equation}
V_G(\Psi) = -m^2\int \frac{|\Psi(r',t)|^2}{|r - r'|}\ \d^3 r'
\label{eq:SNpotential}
\end{equation}
is the potential due to the gravitational self-interaction of the massive particle. Consequently, as emphasized in \cite{Bahrami_2014}, \emph{if} gravity is fundamentally classical and \eref{eq:semiclassicalEFEs} describes how gravity and quantum mechanical matter couple, \emph{then} the SN equation is the fundamental equation of motion for a massive, non-relativistic quantum particle. 

Despite the non-linear nature of the SN equation, the standard probabilistic interpretation of the wavefunction persists. Indeed, the potential due to the gravitational self-interaction \eqref{eq:SNpotential} is a function of $|\Psi|^2$, so under the usual assumptions that $\Psi$ is square-integrable and that its first derivative vanishes at infinity, $\partial_t \int |\Psi(r,t)|^2\, \d^3r = 0$. Consequently, the norm of the wavefunction is preserved under SN evolution \cite{Carlip_2008,Giulini_2023}.

The SN dynamics yield a diversity of novel phenomena that have been extensively studied in the literature \cite{DIOSI1984199,PenroseStateCollapse,Penrose:2014nha,Grossardt_2024,hatifi2023revealingselfgravitysterngerlachhumptydumpty,Sahoo_2023,Delgado_2025}. In the context of solving hard computational problems, the most relevant aspect of the SN dynamics is the non-linearity produced by the gravitational self-interaction. Indeed, it is known that certain non-linear theories of quantum mechanics afford efficient solutions to $\NP$-complete problems (e.g., Weinberg's \cite{WEINBERG1989336ha, Abrams_1998}), so it is plausible that semiclassical gravity could as well.

\subsection{Qubit Dynamics in Weak-Field Semiclassical Gravity}
\label{sec:SNQubitDynamics}

We will now specialize the SN dynamics to the case of a massive and non-relativistic qubit. In particular, consider a non-relativistic spin-1/2 particle of mass $m$ whose overall quantum state is a time-dependent spinor $\ket{\Psi(t)}$. In the position basis, the state of the particle is a two-component wavefunction $\Psi(r,t) = \braket{r}{\Psi(t)}$ in a superposition of spin-up ($\ket{0}$) and spin-down ($\ket{1}$) states,
\begin{equation}
\Psi(r,t) = \psi_\uparrow(r,t) \otimes \alpha \ket{0} + \psi_\downarrow(r,t) \otimes \beta \ket{1},
\label{eq:SpinorSuperpositionUpDown}
\end{equation}
where $|\alpha|^2 + |\beta|^2 = 1$.\footnote{Note, it is generally unclear how entangled states evolve in a non-linear quantum theory because the superposition principle no longer holds. To mend this issue, we follow the prescription of Weinberg \cite{WEINBERG1989336ha} and Abrams and Lloyd \cite{Abrams_1998} by regarding $\{\psi_\uparrow \otimes \ket{0}, \psi_\downarrow \otimes \ket{1}\}$ as a preferred basis on the tensor product Hilbert space, relative to which the action of a non-linear map $S$ on the spin subsystem is specified by acting $S$
independently on each of the computational basis states.}

In the presence of an additional external potential $V_{\text{ext}}(r,t)$ (e.g., a magnetic field like in a Stern--Gerlach experiment \cite{Grossardt_2024,hatifi2023revealingselfgravitysterngerlachhumptydumpty,Sahoo_2023,Delgado_2025}), the time evolution of the wavefunctions $\psi_{\uparrow}$ and $\psi_{\downarrow}$, as predicted by the SN equation \eqref{eq:SNEquation}, is given by
\begin{equation}
i\, \partial_t \psi_{\uparrow\downarrow}(r,t) = \left(-\frac{1}{2m}\nabla^2 + V_G(\Psi) + V_{\text{ext}}(r,t) \right) \psi_{\uparrow\downarrow}(r,t).
\label{eq:SNforQubit}
\end{equation}
Here, the momentum term causes the wavefunction to spread out spatially in time, while the self-gravitational potential $V_G(\Psi)$ introduces a competing effect that inhibits the rate of spreading.\footnote{Incidentally, this gravitational effect was previously conjectured by Diosi and Penrose to be the mechanism behind wavefunction collapse \cite{DIOSI1984199,PenroseStateCollapse,Penrose:2014nha,Moroz1998SphericallySS}.} To see this, note that since $\ket{0}$ and $\ket{1}$ are orthonormal,
\begin{equation}
V_G(\Psi) = |\alpha|^2 V_G(\psi_\uparrow) + |\beta|^2 V_G(\psi_\downarrow).
\label{eq:potentialequation}
\end{equation}
Therefore, as $\psi_{\uparrow\downarrow}$ evolves, it is both gravitationally attracted to itself and $\psi_{\downarrow\uparrow}$, as evidenced by the first and second term in \eref{eq:potentialequation}, respectively, thus combating the spreading \cite{Bahrami_2014}.

Ultimately, these gravitational effects manifest in the dynamics as a non-trivial, initial conditions-dependent phase difference $\Delta \phi_{\SN}$ between the spin-up and spin-down components of the qubit \cite{Grosssardt_2021, Grossardt_2024, Delgado_2025}. In this way, the SN dynamics mimic the dynamics of Weinberg's non-linear theory of quantum mechanics \cite{WEINBERG1989336ha, Delgado_2025}. Intriguingly, in the context of a Stern--Gerlach experiment in which the external potential models a non-uniform magnetic field that is oriented along a fixed axis, one finds that $\Delta \phi_{\SN} \propto (|\alpha|^2 - |\beta|^2) m^2$. Thus, at least in this case, the non-linear effects of the SN equation are maximized for very massive and non-uniform superposition states \cite{Grosssardt_2021}.\footnote{These results are based on a recursive perturbative expansion of the self-interaction potential $V_G$ and several simplifying assumptions. For detailed methodologies and numerical approaches, see \cite{Grosssardt_2021, Grossardt_2024, Delgado_2025, Sahoo_2023, hatifi2023revealingselfgravitysterngerlachhumptydumpty}.}

\section{Solving $\NP$-Complete Problems with Semiclassical Gravity}

In this section, we show how the SN dynamics can be exploited to efficiently solve $\NP$-complete problems. Our idea is largely a SN-rebranding of Abrams and Lloyd's argument that Weinberg's non-linear theory of quantum mechanics can efficiently solve $\NP$-complete problems \cite{Abrams_1998}, as well as Bao, Bouland, and Jordan's \cite{Bao_2016} generalization of Abrams and Lloyd's work to other non-linear theories. To apply their ideas, however, we first need to rephrase a given $\NP$ problem as a state discrimination task, which can then be solved using the SN dynamics.

\subsection{Encoding an $\NP$ Problem into a Qubit}

Given $L \in \NP$, its corresponding algorithm $M$, its corresponding polynomial $p$, and an input $x \in \{0,1\}^*$, let 
\begin{equation}
P_{x} = \left|\left\{y \in \{0,1\}^{p(|x|)} : M(x, y) = 1\right\}\right|
\end{equation}
be the number of proofs that cause $M$ to accept. Evidently, to decide whether $P_{x} = 0$ or $P_{x} > 0$ is sufficient to decide $L$. In \cite{Abrams_1998}, Abrams and Lloyd give a quantum algorithm that can efficiently transform this question to the problem of deciding whether a particular quantum state is $\ket{0}$ or not. Specifically, the Abrams--Lloyd algorithm takes any $L \in \NP$ and prepares the single-qubit quantum state (up to normalization)
\begin{equation}
\frac{2^{p(|x|)} - P_x}{2^{p(|x|)}} \ket{0} + \frac{P_x}{2^{p(|x|)}} \ket{1}.
\label{eq:ALState}
\end{equation}
Therefore, the Abrams--Lloyd algorithm proves that to decide $L$ on a quantum computer, it suffices to be able to reliably decide if \eref{eq:ALState} is $\ket{0}$ or not. Intuitively, this state discrimination task is hardest when $P_x = 1$, i.e., when the language $L$ not only has a proof, but has a \emph{unique} proof. Interestingly, as mentioned in \cite{Bao_2016}, by using ideas from computational complexity theory (namely, the Valiant--Vazirani theorem \cite{Valiant_1986}), one can formalize this intuition and refine the Abrams--Lloyd algorithm so that with sufficiently high probability, deciding $L$ does indeed reduce to this hardest case.
 
\begin{theorem}[Corollary of Abrams and Lloyd \cite{Abrams_1998} and Valiant and Vazirani \cite{Valiant_1986}; see also Bao, Bouland, and Jordan \cite{Bao_2016}]
\label{thm:ALAlgorithm}
For all $L \in \NP$, there exists a polynomial time quantum algorithm $Q_L$ such that for all inputs $x \in \{0,1\}^*$, $Q_L(x)$ outputs the state (up to normalization)
\begin{equation}
\label{eq:languageALstate}
\frac{2^{p(|x|)} - L(x)}{2^{p(|x|)}} \ket{0} + \frac{L(x)}{2^{p(|x|)}}\ket{1}
\end{equation}
with constant probability.
\end{theorem}

In other words, to decide $L \in \NP$, it suffices to be able to reliably distinguish between the single-qubit states $\ket{\varphi_0} = \ket{0}$ and $\ket{\varphi_1} = N\big(\frac{2^{p(|x|)} - 1}{2^{p(|x|)}} \ket{0} + \frac{1}{2^{p(|x|)}}\ket{1}\big)$, where $N$ is the normalization factor. Notice that $\ket{\varphi_1}$ is indeed \eref{eq:ALState} with $P_x = 1$. Now, as $\ket{\varphi_0}$ and $\ket{\varphi_1}$ are exponentially close in $|x|$, one cannot efficiently distinguish them using standard quantum mechanics \cite{Nielsen_Chuang_2010}. However, as we show in the next subsection, one can efficiently distinguish them using the SN dynamics.\footnote{Note that $\ket{\varphi_1}$ is a highly non-uniform superposition state. Thus, per our discussion in Section~\ref{sec:SNQubitDynamics}, the gravitational effects of the SN equation are large for this state, at least in a SN Stern--Gerlach experiment.}

\subsection{The SN Dynamics and the Bao--Bouland--Jordan Algorithm}
\label{sec:ALBBJ}

Let $M(\CC^N)$ be the manifold of normalized pure states on $\CC^N$,\footnote{Therefore, $M(\CC^2)/\mathrm{U}(1) \cong \mathrm{CP}(1)$ is the Bloch sphere, where $\mathrm{U}(1)$ is the unitary group of degree $1$ and $\mathrm{CP}(1)$ is the complex projective space of complex dimension 1.} and equip it with the Fubini--Study metric $d(\ket{a},\ket{b}) = \arccos|\braket{a}{b}|$, which measures the angle between  the states $\ket{a}$ and $\ket{b}$ \cite{Nielsen_Chuang_2010}. Whereas unitary maps are isometries of $M(\CC^N)$, non-unitary maps are not and thus geometrically distort $M(\CC^N)$. An example of this is formalized in the next theorem, which is related to an older result of Mielnik \cite{Mielnik_NonlinearQM,mielnik1985phenomenon}.

\begin{theorem}[Bao, Bouland, and Jordan \cite{Bao_2016}]
Let $S : M(\CC^N) \rightarrow M(\CC^N)$ be a non-unitary diffeomorphism and let $r = \max_{a,b \in M(\CC^N)} \frac{d(S(a), S(b))}{d(a,b)}$. Then, there exists a geodesic $\ell : \RR \rightarrow M(\CC^N)$ such that for all $x,y \in \mathrm{Im}(\ell)$, $\frac{d(S(x), S(y))}{d(x,y)} \geq r$.
\label{thm:BBJThm}
\end{theorem}

In other words, every non-unitary self-diffeomorphism of $M(\CC^N)$ necessarily ``stretches'' or ``magnifies'' some geodesic on $M(\CC^N)$ by a factor of at least $r$. As argued in Abrams and Lloyd \cite{Abrams_1998}, and later generalized by Bao, Bouland, and Jordan \cite{Bao_2016}, this non-isometric stretching can in principle be used to quickly drive two close states far apart, and thus assist in solving hard computational problems.

To see how this works for the SN equation and $\NP$ problems in particular, let $S_{\SN}$ be the map that encodes the net SN evolution of a massive qubit for some fixed amount of time, as discussed in Section~\ref{sec:SNQubitDynamics}. By the discussion in Section~\ref{sec:weakfieldlimit}, $S_\SN$ is a non-linear (and hence non-unitary) diffeomorphism from $M(\CC^2)$ to itself, so $S_{\SN}$ satisfies the conditions of Theorem~\ref{thm:BBJThm}. Therefore, with $r_\SN = \max_{a,b \in M(\CC^2)} \frac{d(S_\SN(a), S_\SN(b))}{d(a,b)}$, there exists a geodesic $\ell_\SN : \RR \rightarrow M(\CC^2)$ such that for all $x,y \in \mathrm{Im}(\ell_\SN)$, $\frac{d(S_\SN(x), S_\SN(y))}{d(x,y)} \geq r_\SN$. In particular,
\begin{equation}
\Delta_\SN = \min_{x,y \in \mathrm{Im}(\ell_\SN)}\frac{d(S_\SN(x), S_\SN(y))}{d(x,y)} \geq r_\SN > 1,
\end{equation}
where the last inequality follows from the fact that $S_\SN$ is non-unitary. Here, $\Delta_\SN$ is the ``magnification factor'' that $S_\SN$ induces on $\ell_\SN$, i.e., it is the minimum amount by which $S_{\SN}$ stretches points on the geodesic $\ell_\SN$.

Now recall $\ket{\varphi_0}$ and $\ket{\varphi_ 1}$ from before, and let $\epsilon = d(\ket{\varphi_0}, \ket{\varphi_1})$ be the distance between them. Evidently, $\epsilon \geq c2^{-q(|x|)}$ for some real constant $c > 0$ and polynomial $q$. Since the length of $\ell_\SN$ is a constant (i.e., it is independent of $|x|$), if $\epsilon$ is greater than the length of $\ell_\SN$, then the states $\ket{\varphi_0}$ and $\ket{\varphi_1}$ can be efficiently distinguished within standard quantum mechanics by preparing a constant number of copies and then measuring them in the computational basis \cite{Nielsen_Chuang_2010}. However, if $\epsilon$ is less than the length of $\ell_\SN$, then this procedure is not guaranteed to work, at least not efficiently. Thus, in this case, apply a single-qubit unitary $U_0$ that maps $\ket{\varphi_0}$ and $\ket{\varphi_1}$ to two points $\ket{\lambda_0^{(0)}}$ and $\ket{\lambda_1^{(0)}}$, respectively, on $\ell_\SN$. Then, apply $S_\SN$ to get states $\ket{\varphi_0^{(1)}}$ and $\ket{\varphi_1^{(1)}}$, respectively, which satisfy $d(\ket{\varphi_0^{(1)}}, \ket{\varphi_1^{(1)}}) \geq \Delta_{\SN} \epsilon$ by Theorem~\ref{thm:BBJThm}. If $\Delta_{\SN}\epsilon$ is larger than the length of $\ell_\SN$, then again use standard techniques to distinguish the two states. However, if $\Delta_\SN\epsilon$ is smaller than the length of $\ell_\SN$, then repeat again: apply the single-qubit unitary $U_1$ that maps $\ket{\varphi_0^{(1)}}$ and $\ket{\varphi_1^{(1)}}$ back onto $\ell_\SN$ and reapply $S_\SN$. This gives states $\ket{\varphi_0^{(2)}}$ and $\ket{\varphi_1^{(2)}}$ separated by distance at least $\Delta_\SN^2 \epsilon$, and so on. By iterating this procedure, the states are separated to a constant distance with just
\begin{equation}
\log_{\Delta_{\SN}}(1/\epsilon) \leq \log_{\Delta_\SN}(2^{q(|x|)} / c) = O(q(|x|))
\end{equation}
applications of $S_\SN$. Since the single-qubit unitaries $U_i$ can be precomputed by knowledge of $\ket{\varphi_0}$, $\ket{\varphi_ 1}$, and how $S_\SN$ acts on $M(\CC^2)$,\footnote{Precomputing might feel like cheating, but in fact it is not. In particular, since $S_\SN$ is a fixed map, and since the initial states $\ket{\lambda_0^{(0)}}$ and $\ket{\lambda_1^{(0)}}$ on $\ell_\SN$ are known, the states $\ket{\varphi_0^{(1)}} = S\ket{\lambda_0^{(0)}}$ and $\ket{\varphi_1^{(1)}} = S\ket{\lambda_1^{(0)}}$ can be explicitly computed. Then, to find $U_1$, it suffices to find two points $\ket{a}$ and $\ket{b}$ on $\ell_\SN$ with the same inner product as $\ket{\varphi_0^{(1)}}$ and $\ket{\varphi_1^{(1)}}$, and then define $U_1$ as any unitary for which $U_1\ket{\varphi_0^{(1)}}=\ket{a}$ and $U_1\ket{\varphi_1^{(1)}}=\ket{b}$. The proof of Theorem~\ref{thm:BBJThm} in \cite{Bao_2016} is constructive and independent of $|x|$, so one can access the points on $\ell_{\SN}$ in constant time and easily find such an $\ket{a}$ and $\ket{b}$. Similar reasoning works for the later steps of the algorithm. Crucially, this precomputation would be intractable had we used the state \eqref{eq:ALState} instead of \eqref{eq:languageALstate}, as we would not know what initial states to evolve.} the SN map $S_\SN$ can be used to distinguish between $\ket{\varphi_0}$ and $\ket{\varphi_1}$ efficiently, and thus it can be used to efficiently decide any language in $\NP$.\footnote{Incidentally, this algorithm can also efficiently solve $\sharp\P$ (``sharp $\P$'') problems \cite{Abrams_1998, Bao_2016}, which count the number of proofs of $\NP$ problems. For this reason, $\sharp \P$ problems are even harder than $\NP$ problems \cite{AB09}.}

\section{Discussion}
\label{sec:discussion}

In this paper, we showed that the weak-field dynamics of the semiclassical EFEs can efficiently solve $\NP$-complete problems. In consequence, if gravity is fundamentally classical and if the semiclassical EFEs describe the matter-gravity coupling, then one can efficiently solve $\NP$-complete problems. As this violates the PECTT, we are led to the conclusion that either gravity is not classical, or gravity is classical, but the semiclassical EFEs do not correctly describe the matter-gravity coupling.

However, given the incredibly general nature of Theorem~\ref{thm:BBJThm}, our results suggest that, in fact, \emph{any} consistent, semiclassical, and non-linear matter-gravity coupling will entail an efficient algorithm to solve $\NP$-complete problems and thus violate the PECTT. We therefore take our results not as an argument for an alternative theory of semiclassical gravity, but rather as an argument for the quantization of gravity itself.

Of course, there are many ways that one may try to evade this conclusion, such as by modifying the non-linear structure of the matter-gravity coupling or by simply rejecting the PECTT.\footnote{For example, if the dynamics are such that the density matrix of a quantum state evolves linearly and stochastically (as in Oppenheim's post-quantum theory \cite{Oppenheim_PostQuantumClassicalGravity}), then our conclusions do not hold.} Yet each such move comes at a conceptual cost and requires considerable justification. By contrast, quantizing gravity restores linearity and thus avoids our argument altogether. For example, in the specific context of the weak-field limit of the semiclassical EFEs, quantizing the spacetime metric causes the self-gravitational potential in the SN equation to drop out, thus linearizing the dynamics \cite{Anastopoulos_2014,Giulini_2023}.\footnote{In particular, if the metric perturbation $h_{\mu \nu}$ is quantized, then the matter-gravity coupling $\hat{h}_{\mu \nu} \hat{T}^{\mu \nu}$ in the interaction Hamiltonian leads to linear weak-field dynamics.} Consequently, at least in this case, our argument in Section~\ref{sec:ALBBJ} does not apply, and the PECTT is not obviously violated.

While preparing this manuscript, we became aware of the series of works \cite{Kent_2005,Kent_2021,Fedida_2026} by Kent et al.~concerning hypothetical ``readout devices'' capable of extracting classical information from a quantum system without disturbing it. In \cite{Kent_2021}, it was argued that semiclassical gravity may provide a physically realizable readout mechanism and, moreover, that such devices could in principle implement the original Abrams--Lloyd algorithm in \cite{Abrams_1998} for solving $\NP$-complete problems. In this respect, their framework is closely related to the gravitational mechanism studied here, but our conclusions differ primarily in their interpretation. Specifically, whereas Kent et al.~emphasize the possibility that such dynamics could lead to experimentally accessible computational advantages beyond standard quantum theory, we emphasize that there is a profound tension between such vast computational consequences and the PECTT. Thus, insofar as one regards the PECTT as a fundamental principle of physics, such readout devices cannot exist.

Ultimately, we hope our results inspire a more computational way of thinking about demarcating fundamental physical theories. While gravitational physics and computational complexity theory are rather disparate fields, here we have demonstrated an intriguing consistency check between the two. Specifically, if the PECTT is taken seriously as a principle of physics, then any candidate theory of gravity must respect it. In this way, the PECTT, and computational complexity more broadly, provides not just a constraint, but a guiding principle in the search for a consistent theory of quantum gravity.

\acknowledgments{
The authors are grateful to Eivind J{\o}rstad, Charlie Cummings, Mathew Bub, Thomas Galley, Flaminia Giacomini, and Andr{\'e} Gro{\ss}ardt for discussions related to this project. We are particularly grateful to Ning Bao for explaining the details of the Bao--Bouland--Jordan algorithm in \cite{Bao_2016}. 

Research at Perimeter Institute is supported in part by the Government of Canada through the Department of Innovation, Science and Economic Development Canada and by the Province of Ontario through the Ministry of Colleges and Universities.
}

\bibliography{refs}

\end{document}